%% file: main.tex
\documentclass[conference]{IEEEtran}
\IEEEoverridecommandlockouts
\usepackage{cite}
\usepackage{amsmath,amssymb,amsfonts}
\usepackage{algorithmic}
\usepackage{graphicx}
\usepackage{textcomp}
\usepackage{xcolor}

\def\BibTeX{{\rm B\kern-.05em{\sc i\kern-.025em b}\kern-.08em
    T\kern-.1667em\lower.7ex\hbox{E}\kern-.125emX}}
\begin{document}
\title{Bringing DNS Service to 5G Edge for Reduced Latencies in mMTC Applications \\
}

\author{\IEEEauthorblockN{
Ricardo Harrilal-Parchment\IEEEauthorrefmark{1}, Diana Pineda\IEEEauthorrefmark{1}, Kemal Akkaya\IEEEauthorrefmark{1}, Abdullah Aydeger\IEEEauthorrefmark{2}, and Alexander Perez-Pons\IEEEauthorrefmark{1}} \\

\IEEEauthorblockA{\IEEEauthorrefmark{1}\textit{Dept. of Electrical and Computer Engineering, } \textit{Florida International University}, Miami, FL USA  33174 \\
Email: \{rharr119, dpine033, kakkaya, aperezpo\}@fiu.edu \\
\\
\IEEEauthorrefmark{2}\textit{Dept. of  Electrical Engineering and Computer Science, } \textit{Florida Institute of Technology}, Melbourne, FL USA  32901 \\
Email: \{aaydeger\}@fit.edu}

}

\maketitle

\begin{abstract}

5G brings many improvements to cellular networks in terms of performance, such as lower latency, improved network efficiency, and higher throughput, making it an attractive candidate for many applications. One such domain is industrial applications that may require real-time guarantees to transmit time-critical control messages. Assuming the immense number of devices exchanging data in support of Massive Machine-Type Communications (mMTC) applications, the capability of the cellular infrastructure to handle a large number of real-time transmissions may be inadequate. For such cases, there exists an acute desire to reduce any overheads as much as possible in order to guarantee certain deadlines. One such target is the Domain Name System (DNS) service, for which queries precede almost every new network request. This incorporates additional communication delays based on the response time, which in turn is affected by the proximity of the DNS server. While bringing DNS service to the edge has been touted as a logical solution, its integration with 5G systems is still challenging. This is due to the inability to access the DNS query information at the application layer since the User Equipment (UE) traffic is tunneled through to the core network. To this end, we propose a novel approach that can identify DNS queries at the base stations through Software-Defined Networking (SDN) capabilities. Specifically, we develop an SDN controller which is used to identify and extract DNS queries at the base station and handle the query at the edge without going through the 5G core network. This approach was implemented in a virtualized 5G network, in which we demonstrate that it is feasible and can potentially bring significant performance gains, especially in the case of mMTC applications.

\end{abstract}

\begin{IEEEkeywords}
DNS, 5G, SDN, Edge service, mMTC, real-time communication

\end{IEEEkeywords}

\section{Introduction}
5G is the fifth generation of wireless networking technology, following the previous generations of 2G, 3G, and 4G/LTE. It represents a significant leap forward in terms of speed, capacity, and connectivity, and has the potential to revolutionize the way we communicate and interact with technology with capabilities such as higher speeds, lower latency, increased capacity, and improved reliability. There are many beneficial features associated with 5G, including Enhanced Mobile Broadband (eMBB), Ultra-Reliable Low Latency Communications (URLLC), and Massive Machine-Type Communications (mMTC). Such features position 5G as an attractive option to enable practical solutions supporting many time-critical applications \cite{rao2018impact}. 

Various domains exist under the Industry 4.0 standard, which includes industrial control systems, smart grid, and vehicular networks that require data collection from numerous IoT devices \cite{manavalan2019review}. These environments and applications often impose a restrictive time frame for data acquisition and demand high reliability with real-time application guarantees and feedback. Given that these devices are served through cells, and there will be many other User Equipment (UE) sources that can generate heavy multimedia traffic, the feasibility of deploying a massive number of IoT devices in a limited geographical space needs to be verified. In particular, for mMTC, the real-time requirements of many critical applications need to be evaluated and validated. 


Recently, a significant amount of research has focused on the use of edge computing technologies to optimize network traffic; reducing congestion and enhancing performance \cite{khan2019edge}, although a practical evaluation utilizing actual 5G deployments is often missing. Many challenges arise when attempting to integrate various solutions to the existing standards dictated by 3GPP, and such integration often comes with interoperability issues that may hinder (or break) expected outcomes. For instance, one aspect that has been considered for dealing with real-time delay requirements is to reduce the latency due to Domain Name System (DNS) requests which precede almost every message coming from edge devices, including IoT. Industrial control applications such as ICS/SCADA often dictate extremely low delay requirements (i.e., less than 3ms) \cite{aydeger2019sdn} imposing a very short communications window unless delays due to DNS, security, and other overheads are minimized. Therefore, considering thousands of IoT devices using 5G base stations in a given region, the impact, performance, and scalability of DNS in such environments are becoming increasingly important. 

Given the latency requirements in the order of milliseconds, one can imagine that the time it takes to reach an authoritative DNS server that is far away from a UE device will significantly impact the overall delay, which is critical for mMTC applications. This issue is mitigated through the use of caching resolvers, whose function is to store such responses and serve them up to the requesting nodes without contacting the authoritative servers every single time. It is, therefore, crucial to bring the DNS cache very close to the UE devices in an mMTC scenario. Maintaining the cache on the base stations (e.g., gNB) is possible, but integrating this solution requires a novel approach that will not disrupt existing message flows in 5G communications. 

In this paper, we present the design and implementation of a local DNS server on a testbed implementing a 5G architecture to investigate the feasibility and performance of DNS in this context. The main idea is to extract DNS request from the incoming UE messages at the application layer and direct it to a local DNS server for a faster response. As UE traffic is tunneled via GPRS Tunneling Protocol (GTP) directly between the UE and User Plane Function (UPF), this is not readily available at the Next Generation Node B (gNB). Indeed, this issue has been overlooked in the recent proposals of DNS service at the edge \cite{suzuki2020enhanced}. Unless 5G protocols are re-designed from scratch to enable this feature, deploying these solutions will not be practical. We need an interoperable solution with the existing architecture and mechanisms of 5G systems. 

Therefore, we propose to deploy a software-defined networking (SDN) based approach at the gNB, which can extract DNS queries at the gNB and serve a response locally at the edge. The main motivation for using SDN is to extract the DNS content from the tunneled message and inject responses into the tunnel without interfering with 3GPP-defined specifications for 5G architecture. This allows us to serve these queries locally without returning to the UPF in the 5G Core network. Ideally, the SDN switch will be capable of performing these actions on its own as per defined flow rules, using characteristics such as hardware acceleration and minimizing overhead. Current SDN switches are, however, somewhat limited in terms of autonomous features. So for our implementation, we leveraged an SDN controller to extend the capabilities of the SDN switch, granting more granular control over traffic flows. While the SDN controller does not have to reside with the gNB, in our proof-of-concept, this is desirable to minimize latency between the gNB and controller. To this end, we use virtual functions at the gNB that can locally handle the re-direction of the DNS request to another local process which can serve a response from the local DNS cache. We define the rules at the SDN controller to enable redirection of the traffic.


We implemented this idea in a virtualized 5G testbed \cite{testbed} and conducted preliminary performance tests. The testbed employs a combination of physical and virtual network components, utilizing SDN and Network Function Virtualization (NFV) technologies. Through a series of experiments, we measured the response time of DNS queries and analyzed the impact when an increasing number of DNS requests are sent simultaneously through the same gNB, which we used to emulate an mMTC application running. Compared to a typical architecture, where the DNS server would sit beyond the core network, we show that significant performance gains can be obtained regarding delay reduction. This is key in understanding the feasibility of 5G use for mMTC applications with real-time requirements. 

This paper is organized as follows. The next section provides a summary of the related work. Section III describes the preliminaries and offers background on the use of concepts throughout the paper while also defining the problem. In Section IV, we explain our approach in detail. Section V is dedicated to the implementation of the approach on a testbed and the presentation of the results. Finally, we conclude the paper in Section VI. 

\input{related}

\input{background}

\section{Proposed Approach}
\begin{figure}[ht!]
    \centering
    \includegraphics[keepaspectratio=true,angle=0,width=88mm,trim=4 4 4 4,clip]{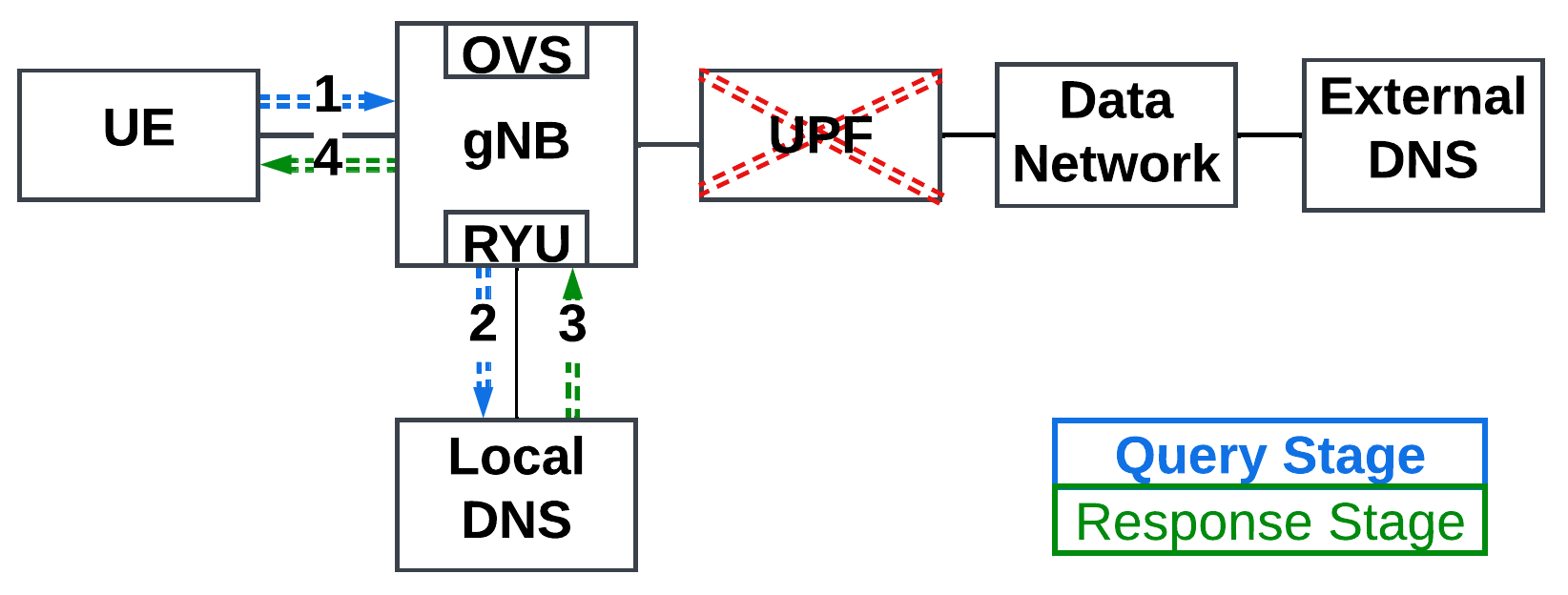}
    \caption{General architecture and approach for processing DNS queries locally.}
    \label{fig:arch}
\end{figure}

Our approach utilizes SDN capabilities to extract DNS queries from the UE, encapsulated in GTP-U in a 5G network environment. Then, we send the DNS query to a local DNS server without relying on the UPF to forward that request to an external DNS server (e.g., Google DNS). For the location of the SDN controller, we opted for the gNB itself, as all the traffic will be passing through it, as illustrated in Fig. \ref{fig:arch}. The rationale behind this decision is to perform faster processing locally at the gNB using network function virtualization capabilities. 

Using the SDN's centralized and southbound application capabilities, we are able to process the GTP-U network traffic in real-time, gather the necessary information from a UE DNS query, send it to the local virtual DNS, and forward the response back to the UE by adequately setting the rules at the SDN controller. Note that this implies that an SDN switch should be deployed at the gNB to divert the packets to the SDN controller. An overall architecture for this approach is shown in Fig. \ref{fig:arch}.

\subsection{SDN Switch Phase}
The first part of our approach consists of creating three flows to the SDN switch on gNB as follows:
 \begin{itemize}
\item A flow to send GTP-U traffic to the SDN controller for handling (outbound UE DNS queries);
\item A flow to send DNS responses from the edge/local DNS server to the controller for handling (inbound DNS responses from the edge);
\item A flow to forward all other traffic normally.
\end{itemize}

 Frames sent to the controller for further analysis are communicated via the OpenFlow protocol and received as {\fontfamily{pcr}\selectfont{packet\_in}} events, and similarly, new frames created by the controller are sent to the switch for delivery are transmitted via OpenFlow {\fontfamily{pcr}\selectfont{packet\_out}} events. 

 It is worth noting that this current solution is suboptimal for practical implementation, as there is additional overhead involved with sending the frame to the controller, having the controller process them, and telling the SDN switch what to do. In an ideal situation, the SDN switch would be capable of processing these frames directly, thus further minimizing the total latency/overhead incurred. This would require more programming capability at the SDN switches, which could be expected to become more widely available in the future. Nonetheless, keeping everything local with gNB would still save us considerable time, as shown in the experiments. 

 \subsection{SDN Controller Phase}
 Once packets containing GTP traffic or from the local DNS server have arrived at the SDN controller, we analyze them further, extract the necessary information, and direct the traffic toward its destination. As shown in Fig. \ref{fig:flowchart}, the general workflow for the controller consists of 3 main stages: 
\begin{itemize}
\item A \textit{learning} stage, where the controller analyzes GTP-U traffic to build a database of connected UEs and their respective tunnels;
\item A \textit{query} stage, where the controller extracts the DNS query from GTP-U traffic and sends it to the edge DNS server;
\item A \textit{response} stage, where the controller injects a DNS response from the edge DNS server into the GTP tunnel between the UPF and the UE.
\end{itemize}

We now detail these stages below: 

\begin{figure}[ht!]
    \centering
    \includegraphics[keepaspectratio=true,angle=0,width=88mm,trim=4 4 4 4,clip]{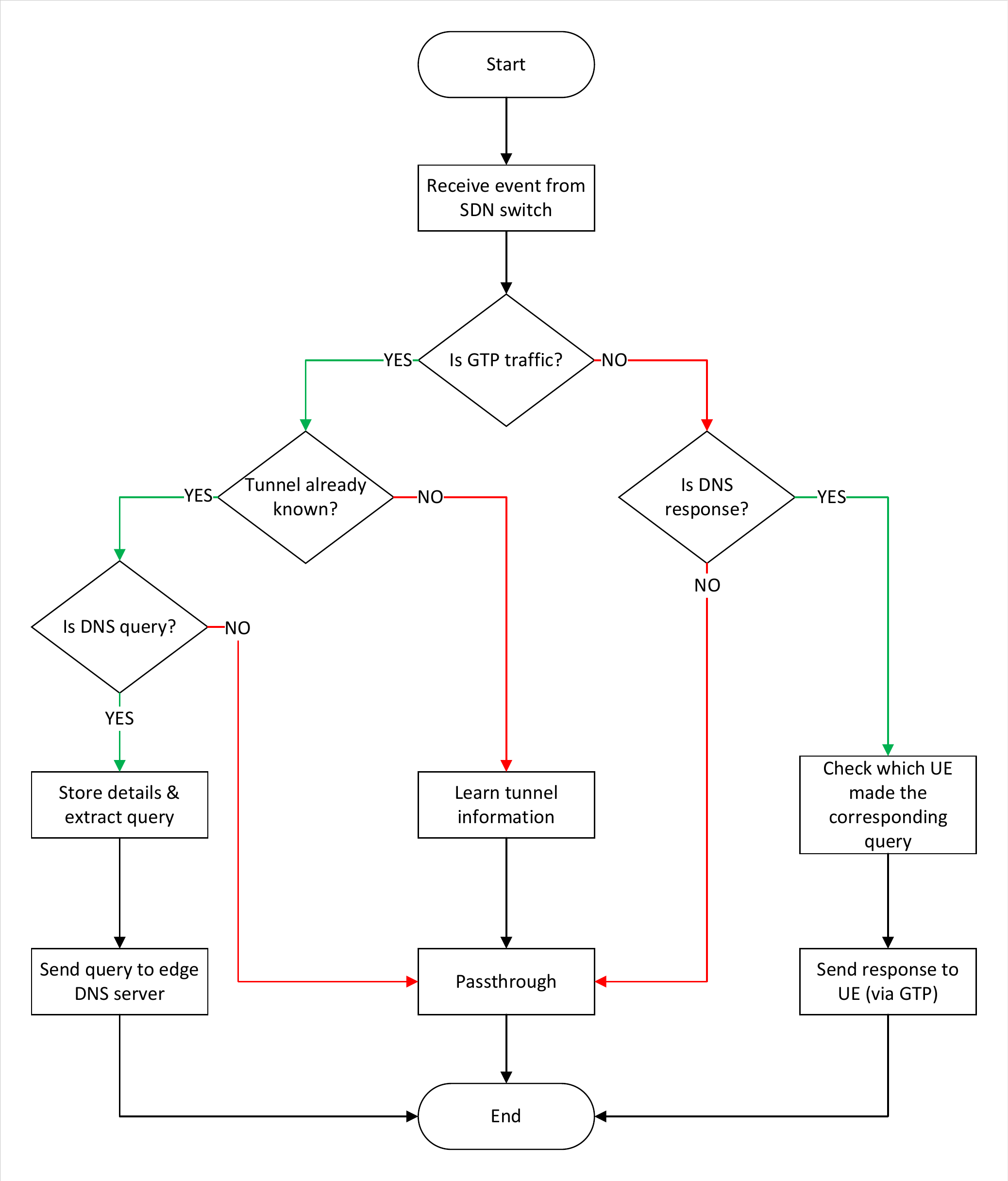}
    \caption{SDN Controller Workflow}
    \label{fig:flowchart}
\end{figure}

\subsubsection{Learning stage}
When a GTP-U packet arrives at the SDN controller, it disassembles the packet into individual layers, consisting of the GTP-U header, GTP-U extension headers, tunneled IPv4 header, subsequent transport layer protocol, and final payload. These layers are analyzed to determine key parameters, such as the UE IP address, GTP Tunnel Endpoint Identifier (TEID), and GTP tunnel type. The controller uses this information to build an internal database of which tunnels are being used by which UEs, and for what type of communication (uplink or downlink).
\subsubsection{Query Stage}
After gathering sufficient data to identify the downlink tunnel for a specific UE, if the controller subsequently identifies a DNS query originating from that UE in the uplink tunnel, it extracts the query and saves critical parameters, such as the requesting UE and DNS transaction ID (see the DNS query on the left branch of the flow diagram in Fig. \ref{fig:flowchart}). It then crafts new transport, network, and ethernet layer headers for the query and subsequently instructs the SDN switch to deliver the frame to the edge DNS, as shown in Fig. \ref{fig:arch}. The DNS service receives the query and processes it normally - serving up a cached response if one is available or obtaining and caching an authoritative one if not.

\begin{figure*}[t!]
    \centering
    \includegraphics[keepaspectratio=true,angle=0,width=185mm]{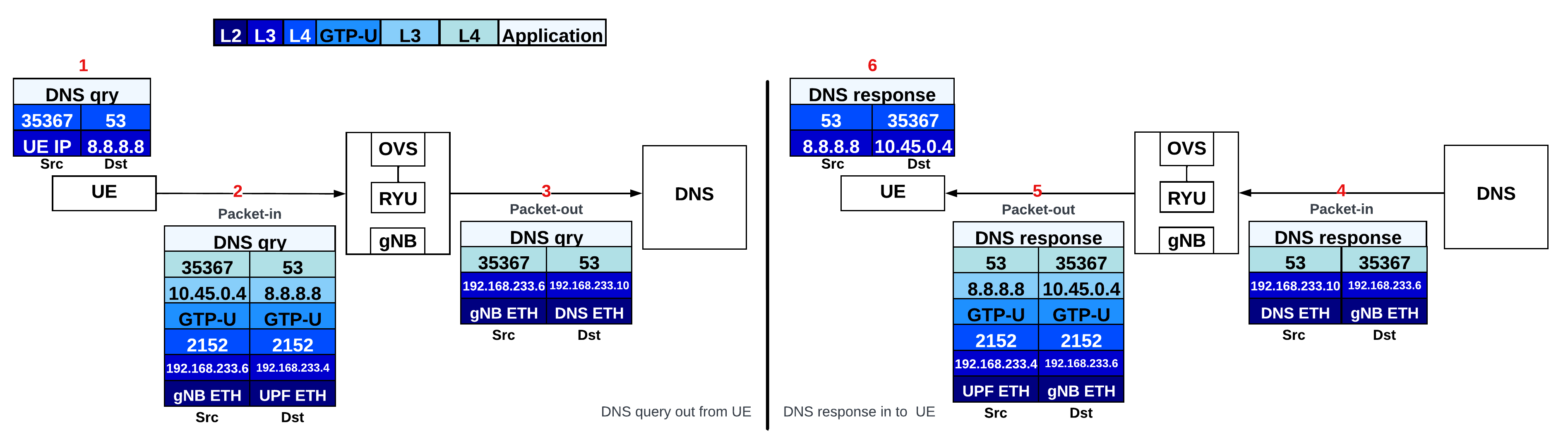}
    \caption{Implementation of Local DNS - Query and Response.}

    \label{fig:workflow}
\end{figure*}

\subsubsection{Response Stage}

  When the controller receives the DNS response, it similarly extracts and matches the response to the stored query (see the DNS response on the right branch of the flow diagram in Fig. \ref{fig:flowchart}). After that, it can leverage the information in its database to encapsulate the response towards a specific UE properly. The controller then crafts new transport, network, and GTP-U with associated extension headers and an additional set of transport, network, and ethernet layer headers for the response. Finally, the controller instructs the SDN switch to deliver the frame to the gNB, which delivers the message to the UE.

\input{experiments}

\section{Conclusion}
In this paper, we propose an integration approach to bring DNS service to the edge for 5G networks. Specifically, we implemented the capability at the base-station by utilizing SDN capabilities. The results of our experiments demonstrated that an SDN-assisted approach to providing DNS services at the edge is feasible in 5G networks. The latency results indicate that this approach can potentially optimize traffic flows, reducing congestion and improving bandwidth utilization throughout the network. These findings have important implications for the design and deployment of future 5G networks, highlighting the potential benefits of incorporating DNS as a key component of the network infrastructure.

\section*{Acknowledgment}
This research was funded by a National Centers of Academic Excellence in Cybersecurity grant (H98230-21-1-0324 (NCAE-C-002-2021)), which is part of the US National Security Agency and National Science Foundation under the grant \#2147196. 

\bibliographystyle{IEEEtran}
\bibliography{master}
\end{document}

%% file: related.tex
\section{Related Work}
This section summarizes other research papers that investigate DNS integration and placement within 5G infrastructure.

NFV and SDN have been some of the main enablers of the 5G networks \cite{yousaf2017nfv}. Specifically, NFV has been utilized in 5G for various reasons such as network slicing \cite{ordonez2017network}, Quality of Service (QoS) \cite{qu2020dynamic}, and security \cite{madi2021nfv}. One of the critical uses of the NFV within the 5G concept would be to run a DNS server as a virtual network function \cite{nesary2022vdns}. Researchers have proposed DNS servers, including destination IP address, 5-tuple packet headers, and domain name, which do not exist in any commercially available Multi-access edge computing for local breakout conditions \cite{kao20215g} in 5G. However, it has not been explained or implemented how and where exactly it would be feasible to have DNS servers. 
In \cite{yucache}, authors propose Top Level DNS (TLD) server to be placed in a local Chinese Mobile Network Provider. This work aims to improve mobile connections in China, which requires DNS resolution, and it does not provide a generic framework for academia. Furthermore, it does not seem to be feasible to employ TLD in every mobile network provider, yet alone within the base stations. In another relevant paper, authors propose DNS servers to support Content Delivery Networks (CDN) speed \cite{hsu2020dns} requirements. While their ideas are similar, the authors in \cite{hsu2020dns} have employed their approaches to Mobile Edge Computing (MEC) without considering 5G network infrastructure and flexibility. In the recent white paper published by European Telecommunications Standards Institute (ETSI) \cite{suzuki2020enhanced}, researchers propose and provide details of DNS servers to every edge at the MEC. Yet, their work would require a real-world implementation of the proposed ideas and proper integration into the 5G networks. 

As summarized, there are a few works that have recently been proposed which are similar to our paper. However, none of these works have successfully integrated recursive DNS within the gNB in a real 5G testbed. We propose a solution to enable recursive DNS server placement within the gNB and generate DNS responses locally for an actual testbed implementation. In addition to the real testbed implementation, we also generated network traffic to empirically evaluate the performance of our environment, comparing it with the benchmark DNS packet exchanges in the 5G networks scenario. To our knowledge, this is the first work that has been able to combine these technologies to develop an operational framework for other researchers to utilize.

%% file: background.tex
\section{Preliminaries}
In this section, we provide some background information for a few topics that are important to understand our proposed approach.

\subsection{DNS}
DNS is the system that maps human-readable domain names (e.g., www.mywebsite.com) to IP addresses (e.g., 192.168.0.1). It works by having a hierarchical structure of servers, where each server is responsible for a portion of the domain namespace. When a client wants to resolve a domain name, it sends a query to a DNS resolver, which then contacts authoritative servers from the top of the hierarchy down until it finds the IP address for the requested domain name. 
To this end, a stub resolver, which runs on a network node, searches its local DNS cache and, if needed, generates and forwards the DNS query to a recursive resolver. A recursive resolver, usually run by the Internet Service Provider (ISP), provides address resolution service and can officiate a level of control over the traffic of its client base. 
If the requested domain name is not found in the DNS recursive resolver's cache, a DNS query is forwarded to one of the 13 Root servers \cite{root_servers}. The Root server provides the address of TLD name servers in which the location of the information about the domain name server will be obtained. As the last step, the DNS query will be sent to the Authoritative name server for the requested domain, where the DNS response will be generated and sent back to the stub resolver eventually.

In 5G networks, the DNS query from a UE is tunneled through the base station directly to the UPF, which then routes the query to the target DNS server. Once received by the DNS server, if the requested address is not in the cache, the query response time varies depending on the location of the authoritative name servers of the requested domain. Even if the network core may employ a DNS cache locally at the UPF, this still requires access to UPF, which is typically far away from the UE devices. 

\subsection{5G User Plane Handling}

\begin{figure}[ht!]
    \centering
    \includegraphics[width=\linewidth]{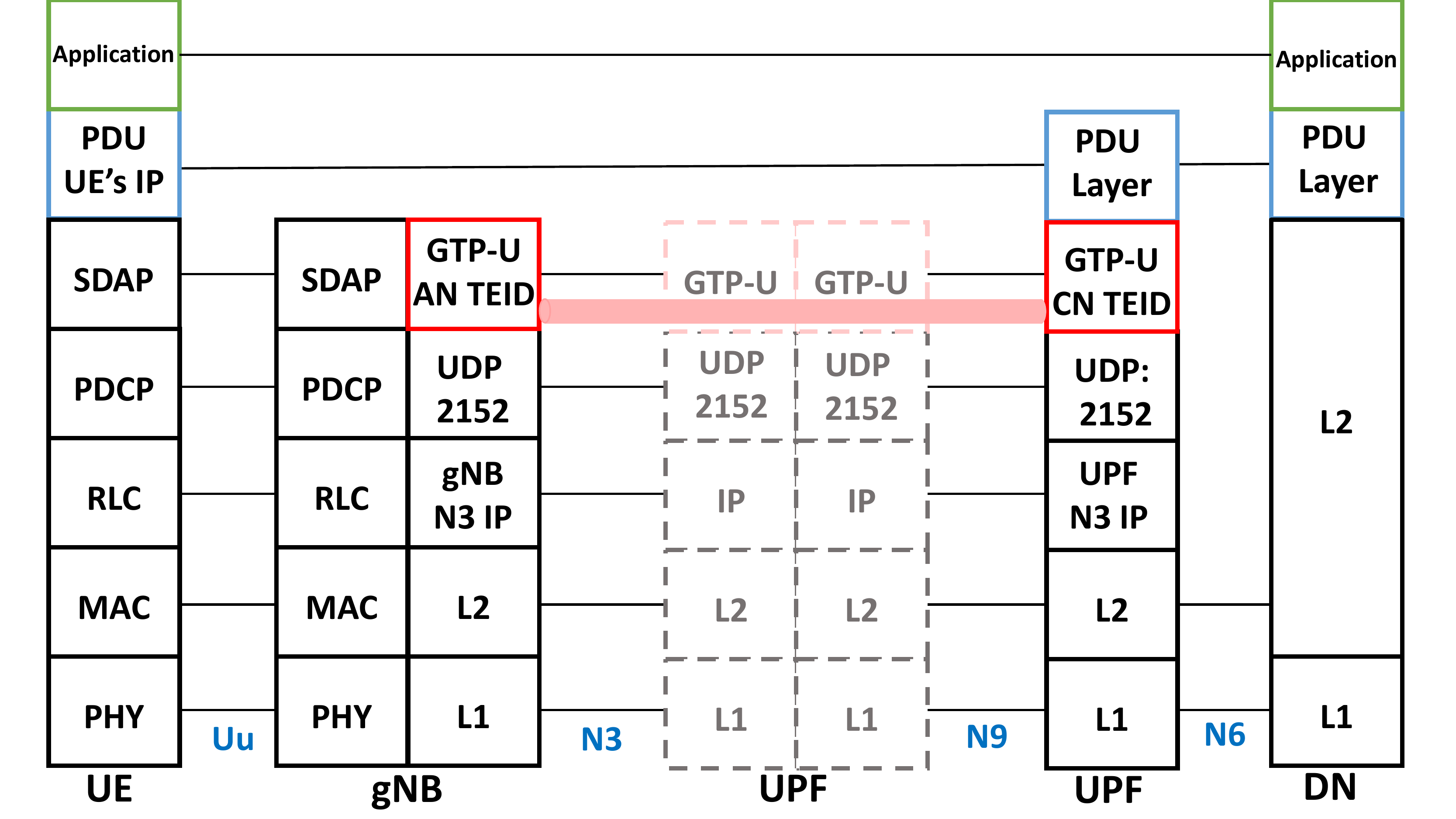}
    \caption{User Plane Protocol Stack for 5G.}
    \label{fig:protocolstack}
\end{figure}

In 5G networks, when a UE IoT device makes a DNS query, it will be encapsulated within an IP packet and sent to the DNS resolver. Therefore, understanding the underlying components of 5G networks, especially the User Plane (UP), is necessary. In this case, the UP domain carries and delivers the data between the UE and the Data Network (DN), and consists of several connections between network functions to allow its primary function. The initial UP connection is between the UE and the gNB, which allows connectivity between mobile devices and the 5G network. The next connection is from the gNB to the UPF in the core network over the N3 reference point.  It is possible that it may involve additional hops between UPFs in the core network using the N9 reference point.  The last connection is between the UPF and the DN across the N6 reference point, as shown in Fig. \ref{fig:protocolstack}. Between the N3 and N9 reference points, the User Plane data is carried in the GTP - User Plane (GTP-U) tunnel. In 5G networks, a reference point indicates the logical point of connection between network functions, and guarantees interoperability and consistency between different mobile network generations \cite{rommer20195g}. 

\subsection{Problem Description}
As mentioned, any data coming from a UE device is encapsulated in a GTP tunnel. Therefore, until the data packet arrives at the UPF, it is difficult to extract the inner payload to process the packets at any location before the 5G core. This is the challenge when it comes to implementing an edge-based DNS resolver to handle DNS queries. While we intend to deploy a DNS resolver/cache that can sit either at the gNB or in between the gNB and core network, there is no normal way that we can detect DNS queries from the packet headers and port numbers on the surface to stop it moving before it arrives the UPF. The problem is detecting that a packet coming from a UE is a DNS query and needs to be directed to a local DNS resolver.

%% file: experiments.tex
\section{Implementation and Testing}
\subsection{Experiment environment}
The proposed methodology has been generated and tested in a 5G-SDN open-source testbed based on the approach found in the documentation provided in \cite{testbed}. This testbed comprises several VMs containing the Control Plane, User Plane, gNB, and UE to simulate the 5G networks and a local DNS server. This environment has been installed in a Dell Precision Workstation with VMware ESXi as the hypervisor. The open-source projects that simulate the 5G networks and DNS edge environment are Open5GS as the Core Network, UERANSIM for the 5G UE and gNB, and Dnsmasq as the DNS caching server—in addition to OpenvSwitch \cite{ovs} as the SDN switch and Ryu \cite{ryusdn} as the SDN controller. Within the ESXi environment, we created seven VMs, including a Control Plane VM, User Plane VM, gNB, UE, OpenvSwitch, Ryu, and DNS server. The VMs related to the 5G network environment are connected to the Openvswitch, which allows them to communicate with each other and with the outside network, as it has been managed by the Ryu controller to have a flexible and scalable SDN environment. Additionally, the gNB VM also has an underlying Openvswitch instance to be able to intercept frames before they enter the 5G RAN and extend SDN capabilities to the gNB, thus completely containing our edge functions at the gNB.

\subsection{Experiments \& Results}
For our experiments, we created a shell script to generate a very large number of DNS queries, measure their individual response time, and calculate the average response time. We incorporated queries to various Top-Level Domains (TLDs) such as .com, .net, .us, .uk, and others, though this was not expected to have a significant impact on the average as the response would be cached after the first. Open5GS includes the \textit{nr-binder} tool, which can be used to run just about any application from a UE device, including our DNS testing script. We then ran several different test scenarios as described below:
\begin{itemize}
\item Queries directed to Google DNS (8.8.8.8) with normal traffic flow
\item Queries directed to OpenDNS (208.67.222.222) with normal traffic flow
\item Queries directed to Verisign DNS (64.6.64.6) with normal traffic flow
\item All DNS queries intercepted \& redirected by our SDN controller to the Edge DNS server
\end{itemize}

The results of our experiments can be seen in Table \ref{tab:avgresp} below:


\begin{table}[htb]
    \renewcommand{\arraystretch}{1.5}
    \centering
    \caption{Average DNS query response times (ms)}
    \label{tab:avgresp}
\begin{tabular}{l|llll|}
\cline{2-5}
                                          & \multicolumn{4}{c|}{\textbf{Number of Queries}}                                                                            \\ \hline
\multicolumn{1}{|l|}{\textbf{DNS Server}} & \multicolumn{1}{l|}{\textbf{10}} & \multicolumn{1}{l|}{\textbf{100}} & \multicolumn{1}{l|}{\textbf{1000}} & \textbf{10000} \\ \hline
\multicolumn{1}{|l|}{Google}              & \multicolumn{1}{l|}{24}          & \multicolumn{1}{l|}{22.32}        & \multicolumn{1}{l|}{21.05}         & 20.12          \\ \hline
\multicolumn{1}{|l|}{OpenDNS}             & \multicolumn{1}{l|}{45.60}       & \multicolumn{1}{l|}{44.56}        & \multicolumn{1}{l|}{35.58}         & 32.30          \\ \hline
\multicolumn{1}{|l|}{Verisign}            & \multicolumn{1}{l|}{52.80}       & \multicolumn{1}{l|}{37.44}        & \multicolumn{1}{l|}{29.40}         & 28.89          \\ \hline
\multicolumn{1}{|l|}{Our Approach}        & \multicolumn{1}{l|}{35.20}       & \multicolumn{1}{l|}{17.60}        & \multicolumn{1}{l|}{17.27}         & 17.26          \\ \hline
\end{tabular}
\end{table}


\subsection{Results Analysis}



From the results, Google DNS seems to be the fastest in both of the public DNS servers we tried, both initially and overall. This is likely due to proximity and how distributed Google's infrastructure is. This was supported by subsequent traceroute outputs, reflecting 8 hops from our testbed to Google DNS and a minimum of 13 for the other two servers. As can be seen, with a lower number of queries, the impact of a lookup not being in the cache is much more significant. There are strategies for reducing the initial delay, such as those discussed in \cite{yucache} through the use of the Centralized Zone Data Service (CZDS) \cite{czds} provided by ICANN/IANA. Comparing performance over time, however, shows once the results have been cached, our experiment yields improved performance compared to other scenarios, with the average response time being close to 20\% faster than Google DNS, which was the next quickest. In all cases, the average latency reduces as the DNS information is stored in the cache after they are fetched in the first responses. This helps to offer much faster responses for later queries.  

As our solution also requires communication between the SDN switch and the controller, it also stands to reason that even faster response times could be obtained if it were possible to remove the controller and execute the workflow entirely within the SDN switch. Eliminating the controller and handling the traffic entirely within the switch would bring the overall process closer to wire speed, as there is no need to relay information back and forth between the controller and switch.

Another aspect to consider in this experiment is the number of DNS requests tested. It is difficult to test a large number of DNS requests against public DNS servers in a real-world setting, as most providers rate limit queries and have policies in place to prevent abuse and ensure fair usage of their resources. DNS has a high potential for abuse and is often leveraged in Denial-of-Service attacks, and sending too many queries within a short space of time may likely get your IP temporarily blocked. We attempted to test with even higher numbers but found that we would eventually stop getting responses though we could see the query traffic leaving our testbed, suggesting that responses were being suppressed by the providers. With the large number of devices present in a real-life mMTC setting though, such levels of DNS queries can be expected and although the corresponding response data is not reported in this paper, our approach in such a scenario is expected be much more impactful.


%% file: main.bbl
\begin{thebibliography}{10}
\providecommand{\url}[1]{#1}
\csname url@samestyle\endcsname
\providecommand{\newblock}{\relax}
\providecommand{\bibinfo}[2]{#2}
\providecommand{\BIBentrySTDinterwordspacing}{\spaceskip=0pt\relax}
\providecommand{\BIBentryALTinterwordstretchfactor}{4}
\providecommand{\BIBentryALTinterwordspacing}{\spaceskip=\fontdimen2\font plus
\BIBentryALTinterwordstretchfactor\fontdimen3\font minus
  \fontdimen4\font\relax}
\providecommand{\BIBforeignlanguage}[2]{{%
\expandafter\ifx\csname l@#1\endcsname\relax
\typeout{** WARNING: IEEEtran.bst: No hyphenation pattern has been}%
\typeout{** loaded for the language `#1'. Using the pattern for}%
\typeout{** the default language instead.}%
\else
\language=\csname l@#1\endcsname
\fi
#2}}
\providecommand{\BIBdecl}{\relax}
\BIBdecl

\bibitem{rao2018impact}
S.~K. Rao and R.~Prasad, ``Impact of 5g technologies on industry 4.0,''
  \emph{Wireless personal communications}, vol. 100, pp. 145--159, 2018.

\bibitem{manavalan2019review}
E.~Manavalan and K.~Jayakrishna, ``A review of internet of things (iot)
  embedded sustainable supply chain for industry 4.0 requirements,''
  \emph{Computers \& Industrial Engineering}, vol. 127, pp. 925--953, 2019.

\bibitem{khan2019edge}
W.~Z. Khan, E.~Ahmed, S.~Hakak, I.~Yaqoob, and A.~Ahmed, ``Edge computing: A
  survey,'' \emph{Future Generation Computer Systems}, vol.~97, pp. 219--235,
  2019.

\bibitem{aydeger2019sdn}
A.~Aydeger, N.~Saputro, K.~Akkaya, and S.~Uluagac, ``Sdn-enabled recovery for
  smart grid teleprotection applications in post-disaster scenarios,''
  \emph{Journal of Network and Computer Applications}, vol. 138, pp. 39--50,
  2019.

\bibitem{suzuki2020enhanced}
M.~Suzuki, T.~Miyasaka, D.~Purkayastha, Y.~Fang, Q.~Huang, J.~Zhu, B.~Burla,
  X.~Tong, D.~Druta, J.~Shen \emph{et~al.}, ``Enhanced dns support towards
  distributed mec environment,'' \emph{ETSI White Paper}, no.~39, 2020.

\bibitem{testbed}
D.~Pineda, R.~Harrilal-Parchment, K.~Akkaya, A.~Ibrahim, and A.~Perez-Pons,
  ``Design and analysis of an open-source sdn-based 5g standalone testbed,'' to
  appear, proceedings of INFOCOM Workshop on Computer and Networking
  Experimental Research using Testbeds (CNERT 2023).

\bibitem{yousaf2017nfv}
F.~Z. Yousaf, M.~Bredel, S.~Schaller, and F.~Schneider, ``Nfv and sdn—key
  technology enablers for 5g networks,'' \emph{IEEE Journal on Selected Areas
  in Communications}, vol.~35, no.~11, pp. 2468--2478, 2017.

\bibitem{ordonez2017network}
J.~Ordonez-Lucena, P.~Ameigeiras, D.~Lopez, J.~J. Ramos-Munoz, J.~Lorca, and
  J.~Folgueira, ``Network slicing for 5g with sdn/nfv: Concepts, architectures,
  and challenges,'' \emph{IEEE Communications Magazine}, vol.~55, no.~5, pp.
  80--87, 2017.

\bibitem{qu2020dynamic}
K.~Qu, W.~Zhuang, Q.~Ye, X.~Shen, X.~Li, and J.~Rao, ``Dynamic flow migration
  for embedded services in sdn/nfv-enabled 5g core networks,'' \emph{IEEE
  Transactions on Communications}, vol.~68, no.~4, pp. 2394--2408, 2020.

\bibitem{madi2021nfv}
T.~Madi, H.~A. Alameddine, M.~Pourzandi, and A.~Boukhtouta, ``Nfv security
  survey in 5g networks: A three-dimensional threat taxonomy,'' \emph{Computer
  Networks}, vol. 197, p. 108288, 2021.

\bibitem{nesary2022vdns}
M.~M. Nesary and A.~Aydeger, ``vdns: Securing dns from amplification attacks,''
  in \emph{2022 IEEE International Black Sea Conference on Communications and
  Networking (BlackSeaCom)}.\hskip 1em plus 0.5em minus 0.4em\relax IEEE, 2022,
  pp. 102--106.

\bibitem{kao20215g}
L.-C. Kao and W.~Liao, ``5g intelligent a+: A pioneer multi-access edge
  computing solution for 5g private networks,'' \emph{IEEE Communications
  Standards Magazine}, vol.~5, no.~1, pp. 78--84, 2021.

\bibitem{yucache}
H.~Yu, Y.~Liu, L.~Duan, S.~Liu, W.~Wang, D.~Liu, Z.~Peng, and D.~Gong, ``Cache
  top-level domain locally: make dns respond quickly in mobile network,'' 10
  2022.

\bibitem{hsu2020dns}
K.-J. Hsu, J.~Choncholas, K.~Bhardwaj, and A.~Gavrilovska, ``Dns does not
  suffice for mec-cdn,'' in \emph{Proceedings of the 19th ACM Workshop on Hot
  Topics in Networks}, 2020, pp. 212--218.

\bibitem{root_servers}
{IANA}, ``Root servers,'' Available at
  \url{https://www.iana.org/domains/root/servers} [Accessed: Mar. 07, 2023].

\bibitem{rommer20195g}
S.~Rommer, P.~Hedman, M.~Olsson, L.~Frid, S.~Sultana, and C.~Mulligan, \emph{5G
  Core Networks: Powering Digitalization}.\hskip 1em plus 0.5em minus
  0.4em\relax Academic Press, 2019.

\bibitem{ovs}
\BIBentryALTinterwordspacing
``Open vswitch,'' 2016. [Online]. Available: \url{http://www.openvswitch.org/}
\BIBentrySTDinterwordspacing

\bibitem{ryusdn}
\BIBentryALTinterwordspacing
``Ryu sdn framework,'' 2017. [Online]. Available:
  \url{https://ryu-sdn.org/index.html}
\BIBentrySTDinterwordspacing

\bibitem{czds}
\BIBentryALTinterwordspacing
``Centralized zone data service,'' 2014. [Online]. Available:
  \url{https://czds.icann.org/home}
\BIBentrySTDinterwordspacing

\end{thebibliography}
